\documentclass[12pt,
aps,preprint
showpacs,nofootinbib,floatfix]{revtex4}

\usepackage{graphicx}
\usepackage{float}

\newcommand{\be}{\begin{equation}}
\newcommand{\ee}{\end{equation}}
\newcommand{\ba}{\begin{eqnarray}}
\newcommand{\ea}{\end{eqnarray}}

\begin{document}
\begin{titlepage}

\begin{flushright}
\vbox{
\begin{tabular}{l}
    Fermilab-PUB-09-293-T\\	
	OUTP-09-12-P\\
\end{tabular}
}
\end{flushright}

\vspace{0.6cm}

\title{$W+3$ jet production at the Tevatron}

\author{R.~Keith Ellis \thanks{
e-mail: ellis@fnal.gov}}
\affiliation{ Fermilab, Batavia, IL 60510, USA }
\author{Kirill Melnikov \thanks{
e-mail:  kirill@phys.hawaii.edu}}
\affiliation{Department of Physics and Astronomy,\\
Johns Hopkins University,
Baltimore, MD, USA}
\author{Giulia Zanderighi  \thanks{
e-mail: g.zanderighi1@physics.ox.ac.uk}}
\affiliation{Rudolf
Peierls Centre for Theoretical Physics, 1 Keble Road, University of
       Oxford, UK}

\begin{abstract}

\vspace{2mm}

We compute the next-to-leading order QCD corrections to the production
of $W$ bosons in association with three jets at the Tevatron in the
leading color approximation, which we define by considering the number
of colors and the number of light flavors as being of the same order
of magnitude. The theoretical uncertainty in the next-to-leading order
prediction for the cross-section is of the order of 15-25 percent
which is a significant improvement compared to the leading order
result.

\end{abstract}

\maketitle

\thispagestyle{empty}
\end{titlepage}

\section{Introduction}

Production of $Z$ and $W$ bosons in association with jets is an
important process in hadron collider physics \cite{nadolsky}. Indeed,
the inclusive production cross-section provides valuable information
about fundamental parameters of the Standard Model as well as parton
distribution functions. When the inclusive cross-section is split into
components depending on the number of jets, different $V+n$ jet
samples ($V=W,Z$) become backgrounds to a variety of processes that
include top pair production, single top production, Higgs boson
production and many others processes that might appear in extensions
of the Standard Model. To understand these backgrounds, careful
studies of vector boson production in association with jets were
performed at the Tevatron by the CDF and D0
collaborations~\cite{Aaltonen:2007ip,Aaltonen:2007cp,Abazov:2009av}.

The cross-section for $V+n~{\rm jet}$ production depends on how jets
are defined. For typical jet parameters employed by CDF and D0, the
production cross-sections for, say, $W^+ (\to e^+ \nu) + n~{\rm jets}$
range between about thirty picobarns for $n = 1$ and about half of a
picobarn for $n = 3$.  Given the few inverse femtobarns of luminosity
collected at the Tevatron, detailed studies of these cross-sections
are possible. In fact, results on the $W+{\rm jets}$ cross-sections
published by CDF several years ago were based on $320~{\rm pb}^{-1} $
of data.  It is expected that new analyses, based on about one inverse
femtobarn of data, will appear soon.

Tevatron studies of vector boson production in association with
jets~\cite{Aaltonen:2007ip,Aaltonen:2007cp} established proximity, to
within a factor of two, between theoretical predictions based on
leading order (LO) matrix elements merged with parton showers, and
experimental data. While a factor of two sounds like a significant
discrepancy, we note that leading order computations of processes with
a large number of partons always have uncertainties of this order
because different choices of input parameters, such as the
renormalization scale of the strong coupling constant or the
factorization scale in parton distribution functions, lead to large
changes. It is well-known that this problem is significantly reduced
if cross-sections are computed through next-to-leading order (NLO) in
the expansion in the strong coupling constant.

The CDF and D0 collaborations have compared their data with NLO QCD
predictions~\cite{Campbell:2002tg} for $W+n$~jets and $Z+n$~jets, for
$n \le 2$. It turned out that NLO QCD describes data very well. This
is very impressive given that NLO QCD predictions are essentially
parameter-free since unphysical dependencies on renormalization and
factorization scales are reduced to about ten percent. This success of
NLO QCD, combined with the importance of the $V+{\rm jets}$ production
process, suggests that extending NLO QCD predictions to
higher-multiplicity processes may be the best way to describe vector
boson production in association with jets reliably, both at the
Tevatron and the LHC.

Unfortunately, extending NLO QCD computations to $V+3$ and $V+4$ jets
is a difficult task for a number of reasons. For example, in the case
of $V+3$ jets, virtual one-loop amplitudes involve rank five six-point
functions. Real emission corrections require an integration of the
$V(\to l_1 l_2) +6$ parton matrix elements squared over a six-particle
phase-space; the complexity of such a computation approaches the
current frontier for the description of multi-particle processes at
tree level (see e.g. Ref.~\cite{Gleisberg:2008ft}). Clearly, the
complexity of computations required to describe $V+4$ jet production
through NLO QCD is even higher.

In this paper we compute NLO QCD corrections to $W+3~{\rm jet}$
production at the Tevatron.  We extend our previous computation of NLO
QCD corrections to that process \cite{emz} by including all partonic
channels, while working in the leading color approximation defined as
$N_c \sim N_f \gg 1$, where $N_c = 3$ is the number of colors and
$N_f=5$ is the number of massless fermion flavors.  We employ recently
evaluated virtual one-loop amplitudes for $W+5$ partons
\cite{Berger:2008sz,egkmz}. To compute real emission corrections we use the
Catani-Seymour subtraction scheme suitably adapted to deal with the
minimal set of color-ordered amplitudes \cite{emz}.  We perform the
computation within the framework of the MCFM parton level integrator
\cite{mcfm}.

Recently a computation of NLO QCD corrections to $W+3$ jet production
at the Tevatron was reported by Berger {\it et al.} \cite{BH}. That
computation also uses leading color approximation although their
definition of leading color differs from ours. In particular, the
leading color approximation $N_c \gg N_f \gg 1$ 
is employed in Ref.~\cite{BH} only in the
finite parts of virtual corrections, while real emission corrections
are computed with full $N_c$ and $N_f$ dependence.  We also note that
preliminary results for {\it full color} NLO QCD corrections to $W+3$
jets were presented recently \cite{bhtalks}.  Because the computation
that we report here is strictly leading color, it is less accurate
than results reported in \cite{bhtalks}.  Nevertheless, we believe
that our results are worthwhile for two reasons. First, as we will
see, the leading color approximation is sufficiently accurate for the
phenomenology of $W+3$ jets production given the size of experimental
errors. Second, independent cross-checks of these very challenging
computations are important.

The remainder of this paper is organized as follows. In Section II we
present a short summary of theoretical methods employed in the
computation and describe details of the experimental measurement
relevant for a comparison with theory. In Section III we present the
results of our computation of $W+3$ jet process at the Tevatron at
leading and next-to-leading order. We conclude in Section IV.

\section{Setup of the calculation}

\subsection{Theoretical set up}

Our goal is to compute the $W+3$ jet production cross-section through
next-to-leading order in perturbative QCD in the leading color
approximation. We define this approximation as follows.  We consider
the limit $N_c \sim N_f \gg 1$, where $N_c$ is the number of colors,
$N_f$ is the number of massless flavors and we drop terms that are
subleading in $N_c$.
We keep all terms ${\cal O}(N_f/N_c)$. This has the consequence that
both fermion loops in virtual corrections and processes with fermion
pairs in the final state are retained. Fermion loops contribute to the
running of the coupling constant, which is responsible for large scale
variations of the leading order cross-section (see below). Since, for
$N_f = 5$, fermion contributions reduce the QCD beta-function by about
20 percent, fermion loops may be important numerically.  Once the $N_c
\sim N_f \gg 1$ approximation is adopted, all partonic channels that
contribute to hadron-hadron collisions need to be considered.

Any NLO QCD computation requires three ingredients -- one needs to
compute one-loop virtual corrections to the relevant leading order
process, real emission corrections and subtraction terms. Many issues
related to dealing with these three ingredients are standard, at least
as a matter of principle.  In practice, for $W+3$ jet production, the
required one-loop and tree-level computations approach a degree of
complexity where standard tools may not work efficiently. For example,
there are of the order of 1500 one-loop diagrams that contribute to
$W+5$ parton scattering amplitudes. These diagrams involve rank five
six-point functions which are the current frontier in one-loop
computations\footnote{We note that recently the complete calculation
of NLO QCD corrections to $PP\to t\bar t b\bar b$ was performed with
techniques that rely on Feynman diagrams and numerical reduction of
one-loop tensor integrals ~\cite{Bredenstein:2009aj}. That calculation
also involves a comparable number of diagrams, however the tensor
structure is simpler (it involves only up to rank four hexagons) and
the number of subprocesses to be considered is considerably
reduced. Nevertheless, the performance of the code used in that
calculation is impressive both in terms of speed and stability.}.

Real emission corrections require computations of complexity similar
to $W+4$ jets at tree level.  Although no loops are involved in the
latter case, this computation is very challenging because of the
effort required to compute the relevant matrix elements and to
integrate them over high-multiplicity phase-space of the final state
particles. In the next few paragraphs we describe some ideas that we
used to overcome these difficulties.

Our computation of one-loop virtual amplitudes for $W+3$ jets employs
a particular technique called generalized $D$-dimensional
unitarity~\cite{egkm}. It is one of several approaches pursued
currently \cite{Ossola:2007ax,Berger:2008sj, vanHameren:2009dr} which
are based on a connection between one-loop scattering amplitudes and
tree-level amplitudes for complex on-shell
momenta~\cite{Bern:1994zx,Bern:1994zxy,Bern:1998zxyw,Britto:2004nc}.
All amplitudes required for the $W+3$ jets computation are described
in Ref.~\cite{egkmz}.

Our treatment of the real emission corrections is based on the
Catani-Seymour dipole subtraction
formalism~\cite{Catani:1996vz}. However, some modifications of the
formalism are required in our case since we deal with leading color
amplitudes and extensively use symmetry of the final state phase-space
to reduce the number of color-ordered amplitudes that need to be
calculated. Modifications of the subtraction formalism as well as
issues related to our treatment of multi-particle phase-space are
discussed in Ref.~\cite{emz}.

Because we employ the leading color approximation, it is important to
discuss its accuracy. We may get an idea about the quality of the
leading color approximation by studying $W+3$ jets at leading order
and $W+n$ jets, $n \le 2$, at next-to-leading order.

We find that for $W+3$ jet at leading order, the leading color
cross-section exceeds full color cross-sections by about ten
percent. This result seems to hold for various observables and is
independent of the choice of the renormalization and factorization
scales. This independence suggests that rescaling the leading color
production cross-section by a constant factor will lead to an improved
estimate of the cross-section beyond leading order. Therefore, we
define our best approximation to a generic  observable ${\cal O}$ 
computed through NLO
QCD as
\be
\frac{{\rm d} \sigma^{\rm NLO}_{W+3 {\rm jet}}}{{\rm d} {\cal O}}
= {\cal R} \int {\rm d} \sigma^{\rm NLO, LC}_{W+3 {\rm jet}} \;
\delta( {\cal O}(p) - {\cal O})
,\;\;\;
{\cal R} = \frac{\sigma^{\rm LO,FC}_{W+3 {\rm jet}}}{\sigma^{\rm  
LO,LC}_{W+3 {\rm jet}}}.
\label{eq2}
\ee
We call this procedure ``leading color adjustment''. 

The major effect of this procedure on the leading color cross-section
is to rescale the leading order term in Eq.~(\ref{eq2}) making it the
exact leading order full color result.  A similar rescaling of the
next-to-leading order correction in Eq.~(\ref{eq2}) is more
questionable. Nevertheless, as long as the next-to-leading order
correction is not excessively large, this procedure should provide a
sensible approximation to the full color next-to-leading order result.

We can check this assertion by applying the leading color adjustment
 procedure to $W+1$ and $W+2$ jet NLO production cross-sections.
In both cases we find that rescaled leading color cross-sections
agree with full color cross-sections to better than  three percent.

\subsection{Theoretical parameters and implementation
     of experimental cuts}
\label{sec:cuts}
The goal of this Section is to describe precisely what we 
compute theoretically and its relationship to the measurement by the
CDF collaboration. We consider the production of {\it on-shell}
$W^{\pm}$ bosons, that decay into a pair of massless leptons.  We note
that finite width effects are about one percent and that considering
on-shell production tends to {\it overestimate} the cross-section.
We set the CKM matrix to the identity matrix; this {\it reduces} the
$W+3$ jet production cross-section at the Tevatron by about one
percent. All quarks, with the exception of the top quark, are
considered massless. The top quark is considered infinitely heavy and
its contribution is neglected. The mass of the $W$ is taken to be $m_W
= 80.419~{\rm GeV}$. $W$ couplings to fermions are obtained from
$\alpha_{\rm QED}(m_Z) = 1/128.802$ and $\sin^2
\theta_W = 0.230$.  We use CTEQ6L parton distribution functions for
leading order and CTEQ6M for next-to-leading order computations
\cite{Pumplin:2002vw,Nadolsky:2008zw} 
corresponding to $\alpha_s(M_Z) = 0.130$ and $\alpha_s(M_Z) = 0.118$
respectively.  We quote results for three (dynamical) renormalization
and factorization scales $\mu = [\mu_0/2, \mu_0, 2 \mu_0]$, where
$\mu_0 = \sqrt{p_{\perp,W}^2 + m_W^2}$ and  $p_{\perp,W}$ is the
transverse momentum of the  $W$ boson.  We choose these input parameters
to stay maximally close to the choices made in Ref.~\cite{BH} in order to
facilitate a comparison between the two results, to the extent
possible.

In the following we present results for the cuts employed in the
analysis by the CDF collaboration~\cite{Aaltonen:2007ip}.  We require
that the transverse energy and pseudorapidity of the jets satisfy
$E_{\perp,j} > 20~{\rm GeV}$ and $|\eta_j| < 2$ and employ the
following restrictions on lepton transverse momentum, missing
transverse energy, lepton rapidity and transverse invariant mass
$p_\perp^{e} > 20~{\rm GeV}$, $\slash\!\!\!\!E_{\perp} > 30~{\rm
GeV}$, $|\eta_e| < 1.1$, $M_{\perp}^W > 20~{\rm GeV}$. We do not apply
an isolation cut on the leptons since it is removed by the acceptance
correction applied to the experimental results.

To define jets, the CDF collaboration uses the JETCLU cone algorithm
with $R=0.4$ and merging parameter $f=0.75$. Since this algorithm is
not infrared safe, it can not be used in a next-to-leading order
calculation of $W+3$ jets.  As discussed in the next Section, for the
computations reported in this paper we choose to use a somewhat
related but infrared safe seedless cone (SIScone)
algorithm~\cite{siscone} and the anti-$k_\perp$ jet
algorithm~\cite{Cacciari:2008gp}.

\section{Leading order results}

In this Section, we summarize leading order results for total
cross-sections. We note that there is a subtlety associated with the
way CDF presents their results.  While jets are required to have
transverse energy in excess of $20~{\rm GeV}$, the total cross-section
for $W+3$ jet production is quoted with an {\it additional restriction
} -- the transverse energy of the third hardest jet should satisfy
$E_{\perp}^{3{\rm rd~jet}} > 25~{\rm GeV}$.  The CDF measurement
yields the inclusive cross-section
\begin{equation} 
\sigma^{W+\ge 3j}_{E_\perp^{3{\rm rd~jet}}>
25~{\rm GeV}} = (0.84 \pm 0.10\, ({\rm stat.}) \pm 0.21\, ({\rm sys.})
\pm 0.05\, ({\rm lum.}))~{\rm pb}.
\label{eq:CDF}
\end{equation}
Note that this result includes both $W^+(e^+ \nu)$ and $W^-(e^- \bar
\nu)$ production which, given the charge symmetry of the initial state
at the Tevatron, simply doubles the cross-section for fixed $W$
charge.

We now discuss the choice of the jet algorithm in more detail.  As we
already mentioned, the CDF collaboration uses the infrared unsafe
JETCLU jet algorithm. We remind the reader that infrared unsafety
arises because one searches for stable cones around few fixed points
(seeds) in the $\eta-\phi$ plane and one might miss a stable cone. It
is often argued \cite{Ellis:2007ib} that this is mainly an issue of
perturbative calculations, which involve only few particles, and
therefore few seeds. In a true experimental environment, with many
soft emissions giving rise to many seeds, stable cones are rarely
missed and the difference between infrared safe and unsafe jet
algorithms is claimed to be small~\cite{Ellis:2007ib}.

However, even if such claims are true, our inability to use infrared
unsafe jet algorithms in theoretical computations makes it necessary
to choose a jet algorithm which on one hand is infrared safe and, on
the other hand, is close to JETCLU in phenomenological applications.
To facilitate such a choice, in Table~\ref{jetalg} we compare leading
order results for $W+3$ jet production cross-sections for two
different jet algorithms, SISCone and anti-$k_\perp$. We observe that
different jet algorithms for identical values of $R$ lead to different
results -- not an unexpected conclusion.

If we consider jet algorithms for fixed value of $R =0.4$, we find
that anti-$k_\perp$ jet algorithm provides results that are closest to
JETCLU. The SIScone algorithm does not do particularly well -- the
difference between SIScone and JETCLU with the same $R$ is about
twenty percent.
Perturbative studies of jet substructure~\cite{gpspriv} suggest that
the jet algorithm closest to CDF's JETCLU jet algorithm is
anti-$k_\perp$. Therefore, it appears that anti-$k_\perp$ algorithm
should be chosen for our calculation of $W+3$ jet production
cross-section at the Tevatron.

\begin{tiny}
\begin{table}[th]
\begin{center}
\begin{tabular}{| c | c | c | c |}
\hline
Algorithm & $R$ & $E^{\rm jet}_\perp > 20~{\rm GeV}$  & $E^
{\rm 3rd jet}_\perp > 25~{\rm GeV}$ \\ \hline\hline
JETCLU  & 0.4   &  $1.845(2)^{+1.101(3)}_{-0.634(2)}$   & $1.008(1)^
{+0.614(2)}_{-0.352(1)}$ \\ \hline
SIScone  &  0.4  & $1.470(1)^{+0.765(1)}_{-0.560(1)}$ & $0.805(1)^
{+0.493(1)}_{-0.281(1)}$ \\ \hline
anti-$k_\perp$  &  0.4  & $1.850(1)^{+1.105(1)}_{-0.638(1)}$ & $1.010(1)^
{+0.619(1)}_{-0.351(1)}$ \\ \hline
\end{tabular}
\caption{\label{jetalg} Leading order cross-sections in picobarns
for $W+3$ jets at the Tevatron for different jet algorithms. We use
merging parameter $f = 0.75$ for JETCLU and $f = 0.5$ for SISCone.
The renormalization and factorization scales are set to $\mu_0$.  The
upper (lower) value corresponds to setting both scales to $\mu_0/2$
and $2 \mu_0$, respectively. Statistical errors are also
indicated. Other cuts on jets and leptons are described in the text.}
\vspace{-0.1cm}
\end{center}
\end{table}
\end{tiny}

The caveat in this discussion is that since JETCLU is not an infra-red
safe algorithm, the significance of leading order comparisons is
unclear since radiative corrections can be arbitrarily large. Hence,
it is not obvious that the most appropriate jet algorithm for
theoretical calculations is the one which matches the JETCLU leading
order results. To study this question, we perform the next-to-leading
order calculation using both SIScone algorithm 
with $R=0.4$ and $f=0.5$ and the anti-$k_\perp$  algorithm 
with $R=0.4$. The NLO computation with
the SIScone algorithm allows us to compare our results to that of
Ref.~\cite{BH}. A similar computation with the anti-$k_\perp$
algorithm, would, if we had perfect data, tell us whether the
agreement at leading order between JETCLU and anti-$k_\perp$ is
fortuitous.

We now summarize the leading order results for the two algorithms.
Using the three choices of the renormalization and factorization
scales discussed previously, to set upper and lower bounds on the
cross-section variation, we obtain the following result for
leading-color and full-color leading order cross-sections
\begin{eqnarray}
&& \sigma^{W+\ge 3j,\rm LC}_{LO,E_\perp^{3{\rm rd~jet}}> 25~{\rm GeV}} =
0.89^{+0.55}_{-0.31}~{\rm pb},
\;\;\;
\sigma^{W+ \ge 3j,\rm FC}_{LO,E_\perp^{3{\rm rd~jet}}> 25~{\rm GeV}}
= 0.81^{+0.50}_{-0.28}~{\rm pb},~~~{\rm SIScone};
\label{lorescuts_SIScone} \\
&& \sigma^{W+ \ge 3j,\rm LC}_{LO,E_\perp^{3{\rm rd~jet}}> 25~{\rm GeV}} =
1.12^{+0.68}_{-0.39}~{\rm pb},
\;\;\;
\sigma^{W+ \ge 3j,\rm FC}_{LO,E_\perp^{3{\rm rd~jet}}> 25~{\rm GeV}}
= 1.01^{+0.62}_{-0.35}~{\rm pb},~~~{\rm anti}-k_\perp;
\label{lorescuts_ankt}
\end{eqnarray}
In Eqs.~(\ref{lorescuts_SIScone},\ref{lorescuts_ankt}), central values
are for the scale $\mu_0$ and upper (lower) values are for $\mu_0/2$
and $2 \mu_0$, respectively.  We remind the reader that CTEQ6L parton
distribution functions are used in leading order calculations.

The following comments can be made about
Eqs.~(\ref{lorescuts_SIScone},\ref{lorescuts_ankt}). First, as pointed
out in the previous Section, full color results are lower than the
leading color results by about 10 percent; because of that, we will
use ${\cal R} = 0.91$ for both algorithms to rescale NLO leading color
calculations.  Second, it is apparent from
Eqs.(\ref{lorescuts_SIScone},\ref{lorescuts_ankt}) that, in spite of
using a dynamical scale in the leading order computation, the scale
variation of the leading order cross-section is large.
We can quantify it by introducing the following ratio $ \xi^i =
\sigma^{i}(\mu_{\rm max})/\sigma^{i}(\mu_{\rm min}),$
where $i={\rm LO, NLO}$ defines the order at which the cross-section
is computed and $\mu_{\rm max,min}$  are the scales which
give the largest (smallest)  cross-section for the 
three scales  considered.  
We obtain $ \xi^{\rm LO} \approx 2.5.$ for both SIScone and
anti-$k_\perp$ algorithms.  Note that while precise value of $\xi$ is
cut-dependent, the quoted result is typical -- large variations of the
cross-section come from a strong dependence on the renormalization
scale, due to the fact that the cross-section depends on the third
power of the strong coupling constant, $\sigma^{W+3j}
\sim \alpha_s^3$.

\section{Next-to-leading order results}
In this Section, we summarize the next-to-leading order results.
Working in the leading color approximation, we obtain the inclusive
jet cross-section~\footnote{In the case of anti-$k_\perp$, the central
value quoted corresponds to a scale $\mu= \mu_0/2$, while the upper
(lower) bounds correspond to $\mu=\mu_0$ and $\mu=2 \mu_0$,
respectively.}
\begin{eqnarray}
  && \sigma^{W+ \ge 3j, \rm LC}_{NLO,E_\perp^{3{\rm rd~jet}}> 25~{\rm
GeV}} = 1.01^{+0.05}_{-0.17}~{\rm pb},\qquad~{\rm SIScone}, \\
&& \sigma^{W+ \ge 3j, \rm LC}_{NLO,E_\perp^{3{\rm rd~jet}}> 25~{\rm
GeV}} = 1.10^{+0.01}_{-0.13}~{\rm pb},\qquad~{\rm anti-k}_\perp.
\end{eqnarray}
Scaling these results by the tree-level ratio of full-color and
leading color leading order cross-sections, ${\cal R}$, as explained
in the previous Section, we obtain
\begin{eqnarray}
&& \sigma^{W+ \ge 3j}_{NLO,E_\perp^{3{\rm rd~jet}}> 25~{\rm
GeV}}
= 0.91^{+0.05}_{-0.15}~{\rm pb},\qquad~{\rm SIScone},
\label{ourfc} \\
&& \sigma^{W+ \ge 3j}_{NLO,E_\perp^{3{\rm rd~jet}}> 25~{\rm
GeV}} = 1.00^{+0.01}_{-0.12}~{\rm pb},\qquad~{\rm anti-k}_\perp.
\label{ourfc1}
\end{eqnarray}

The next-to-leading computation shows a significant improvement in
stability with respect to changes in the renormalization and
factorization scales.  Calculating the parameter $\xi$, introduced at
the previous Section, we find $\xi^{\rm NLO}_{\rm SIScone} = 1.25$ and
$\xi^{\rm NLO}_{{\rm anti}-k_\perp} = 1.15$ which implies that {\it
overall} uncertainty in the NLO QCD prediction is twenty five percent
or better.  Compared to leading order predictions, the uncertainty is
reduced by at least a factor of four.

We also find that the difference between NLO cross-sections computed
with SIScone and anti-$k_\perp$ is smaller than the difference between
corresponding leading order cross-sections. Nevertheless, the
difference at NLO is about ten percent and therefore not negligible.
Experimental data seems to be closer to SIScone; however, given a
twenty percent uncertainty in data and up to twenty percent
uncertainty in the NLO results, no inconsistency can be claimed.

\begin{figure}[t]
\begin{center}
\includegraphics[angle=0,scale=0.6]{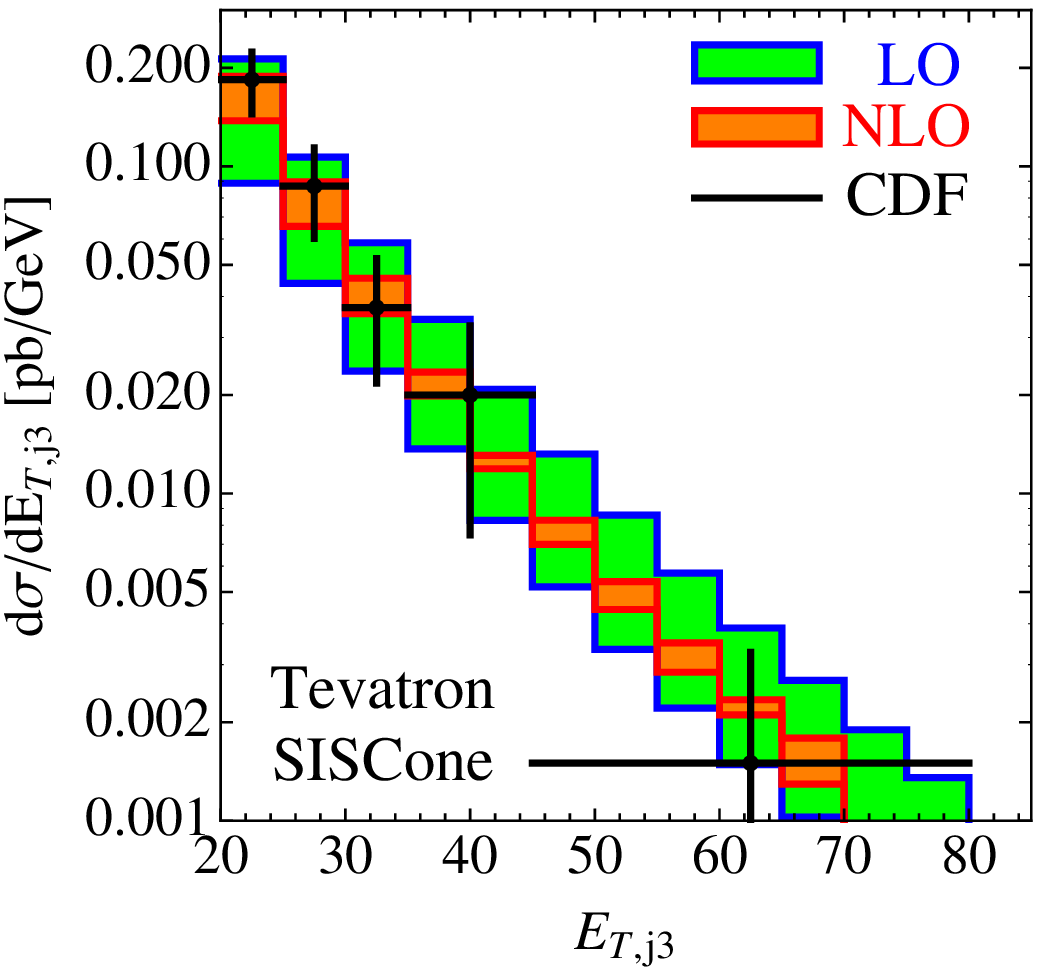}
\hspace{1cm}
\includegraphics[angle=0,scale=0.6]{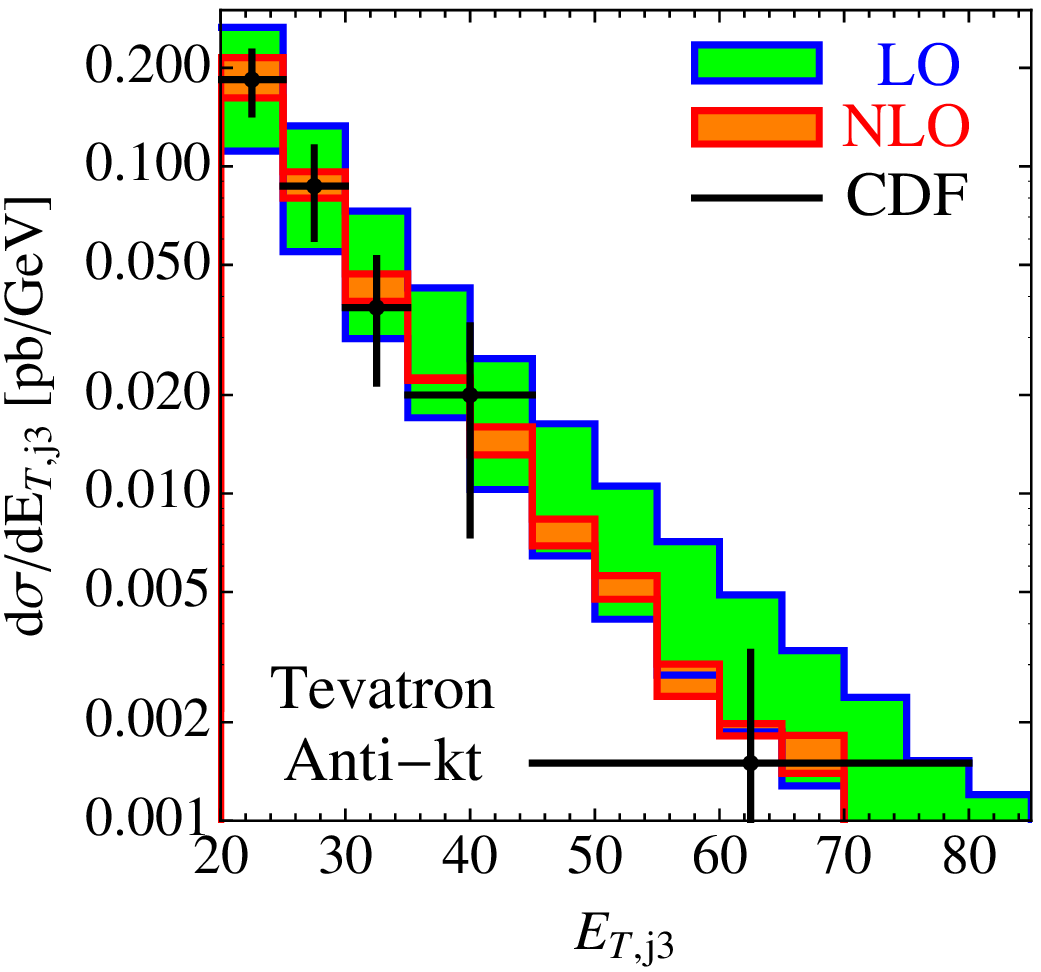}
\caption{The transverse energy distribution of the third hardest jet
for $W+3~{\rm jet}$ inclusive production cross-section at the Tevatron
for SIScone (left) and anti-$k_\perp$ (right) jet algorithms.  All
cuts and parameters relevant for deriving these distributions are
described in the text. Leading color adjustment procedure is applied. 
For experimental points, statistical and
systematic uncertainties are combined in quadrature.  Bands illustrate
scale dependence at leading (green) and next-to-leading order (red).}
\label{fig:1}
\end{center}
\end{figure}

CDF published the transverse energy distribution of the third hardest
jet in $W+3$ jet inclusive production cross-section. In
Fig~\ref{fig:1}, we compare the theoretical prediction for this
distribution at leading and next-to-leading order with experimental
data for the two jet algorithms.  For experimental points, statistical
and systematic uncertainties are combined in quadrature. Theoretical
results are rescaled by ${\cal R} = 0.91$ bin-by-bin, following the
discussion in Section II.  There is reasonable agreement with data,
although uncertainties in current data do not permit high-precision
comparison.

It is interesting to observe that shapes of transverse energy
distributions differ at LO and NLO. Comparing distributions evaluated
at a common scale $\mu_0$, the NLO result exceeds the leading order
cross-section at low values of the third jet transverse energy
$E_\perp^{3 \rm rd}$ and is below the leading order cross-section for
higher values of $E_\perp^{3 \rm rd}$.  Such behavior is consistent
with the expectation that emission of additional QCD partons is
governed by the strong coupling constant evaluated at the transverse
momentum of a ``daughter parton'' defined relative to the direction of
a ``parent parton'' in a QCD branching. When the third hardest jet has
small momentum, the scale $\mu_0 = \sqrt{p_{\perp,W}^2 + m_W^2}$ is
larger than the relative transverse momentum in the branching that
produced this jet; as the result, the leading order computation with
the scale $\mu_0$ underestimates the cross-section.  On the other
hand, when the momentum of the third hardest jet increases, transverse
momenta of the two leading jets start to exceed $\mu_0$; as the
result, the leading order computation with the scale $\mu_0$
overestimates the cross-section.  The change of shape, therefore, can
be attributed to an ``improper'' choice of the coupling constant
renormalization scale in the leading order computation which gets
naturally corrected once one-loop effects are included.  Similar
effects in $W+n$ jet production were recently discussed in
Ref.~\cite{Bauer:2009km} using soft-collinear effective theory.

It is instructive to study the relative importance of the various
subprocesses and how the NLO corrections to them compare.  The
simplest way to split various contributions is into two-quark,
four-quark and six-quark subprocesses where the number of quarks is
the {\it total} number of quarks in $W+3$ jet amplitude (including
those in the loop).  The two-quark and four-quark processes give the 
largest contribution to the cross-section while the six-quark processes
not present at LO are relatively small ($\sim -7\%$). 
We find that NLO QCD corrections affect two-quark and four-quark
processes differently  -- for example, at the reference scale $\mu_0$ 
they increase the two-quark processes by about fifty percent while 
they decrease the four-quark processes by about twenty percent.

There are two lessons that we draw from this observation. First, it
does not seem possible to determine optimal renormalization and
factorization scales for the {\it whole} process by studying NLO QCD
corrections to two- or four-quark processes {\it only}.  Second, we
observe that for the central scale $\mu = \mu_0$, NLO correction to
the {\it total} cross-section is rather modest, about ten
percent. However, this modest correction is the result of a
cancellation between somewhat larger corrections to two-quark and
four-quark channels. This suggests that the leading color adjustment
procedure that we apply may not be very accurate since small
corrections to the adjustment procedure for two- and four-quark
channels {\it separately} may get amplified because of the
cancellation.
Note however that the systematic and luminosity errors on the $W+3$
jet data are currently 25\% and 6\% respectively, see
Eq.~(\ref{eq:CDF}). Given errors of this size, the leading color
approximation seems sufficient for the foreseeable future.

\begin{figure}[t]
\begin{center}
\includegraphics[angle=0,scale=0.59]{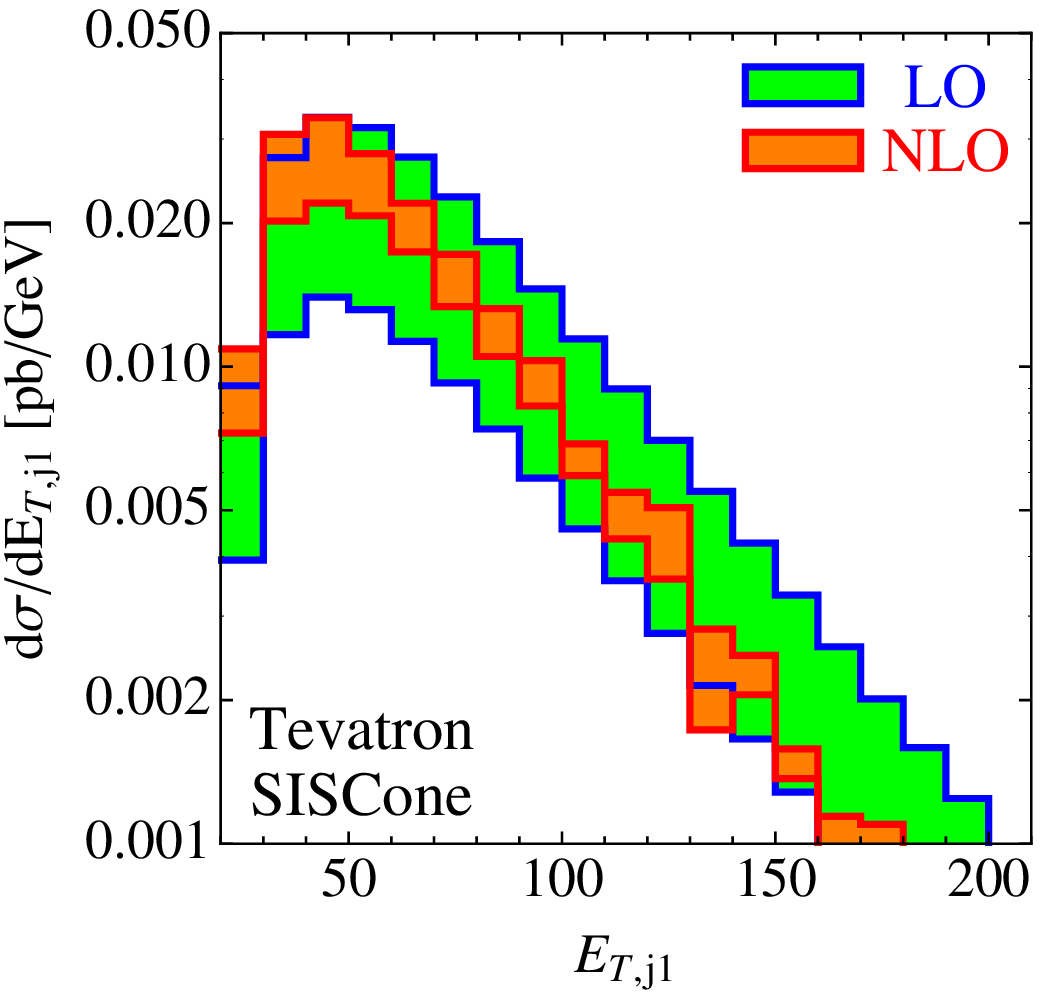}
\hspace{1cm}
\includegraphics[angle=0,scale=0.6]{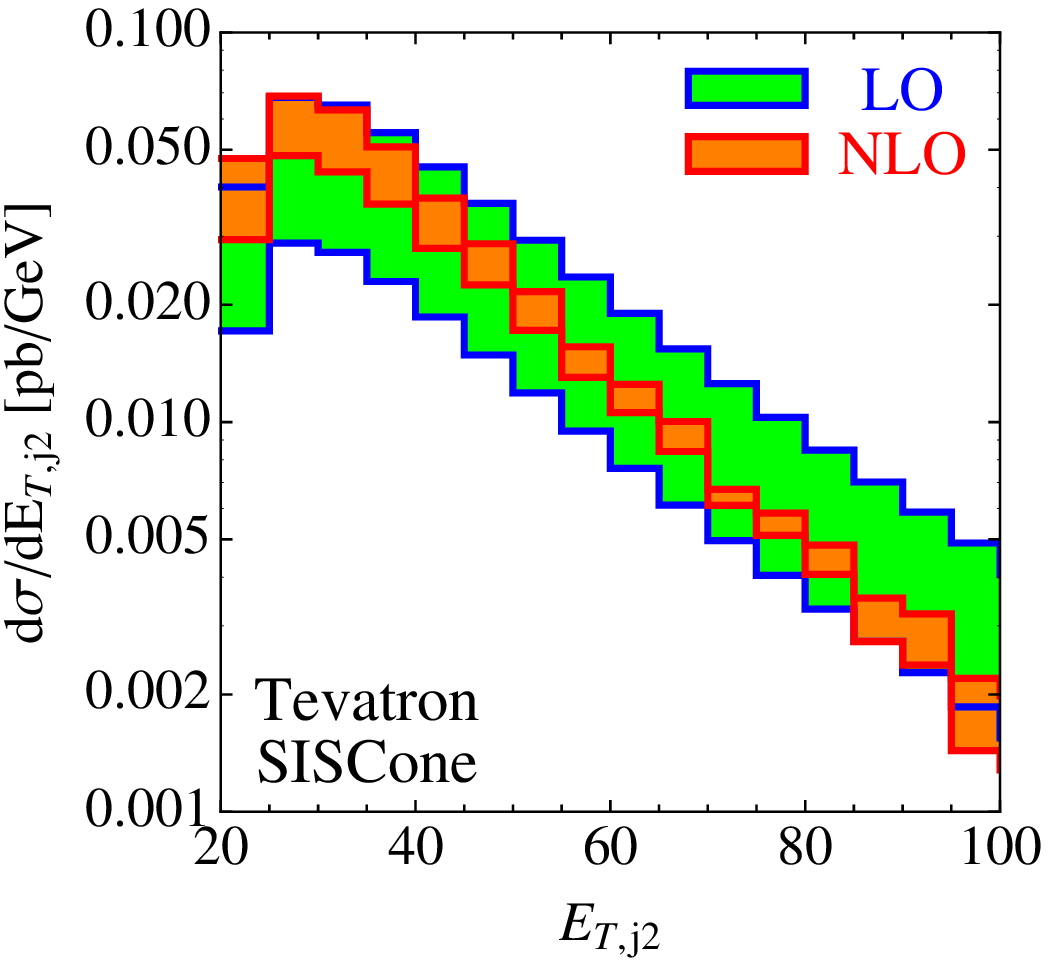}
\caption{The transverse energy distributions of the
hardest (left) and second-to-hardest (right) jet in the $W+3~{\rm
jet}$ inclusive sample using SIScone jet algorithm. All cuts and
parameters are described in the text. Bands illustrate scale
dependence at leading (green) and next-to-leading order (red).
Leading color adjustment is applied.}
\label{fig:2}
\end{center}
\end{figure}

\begin{figure}[t]
\begin{center}
\includegraphics[angle=0,scale=0.59]{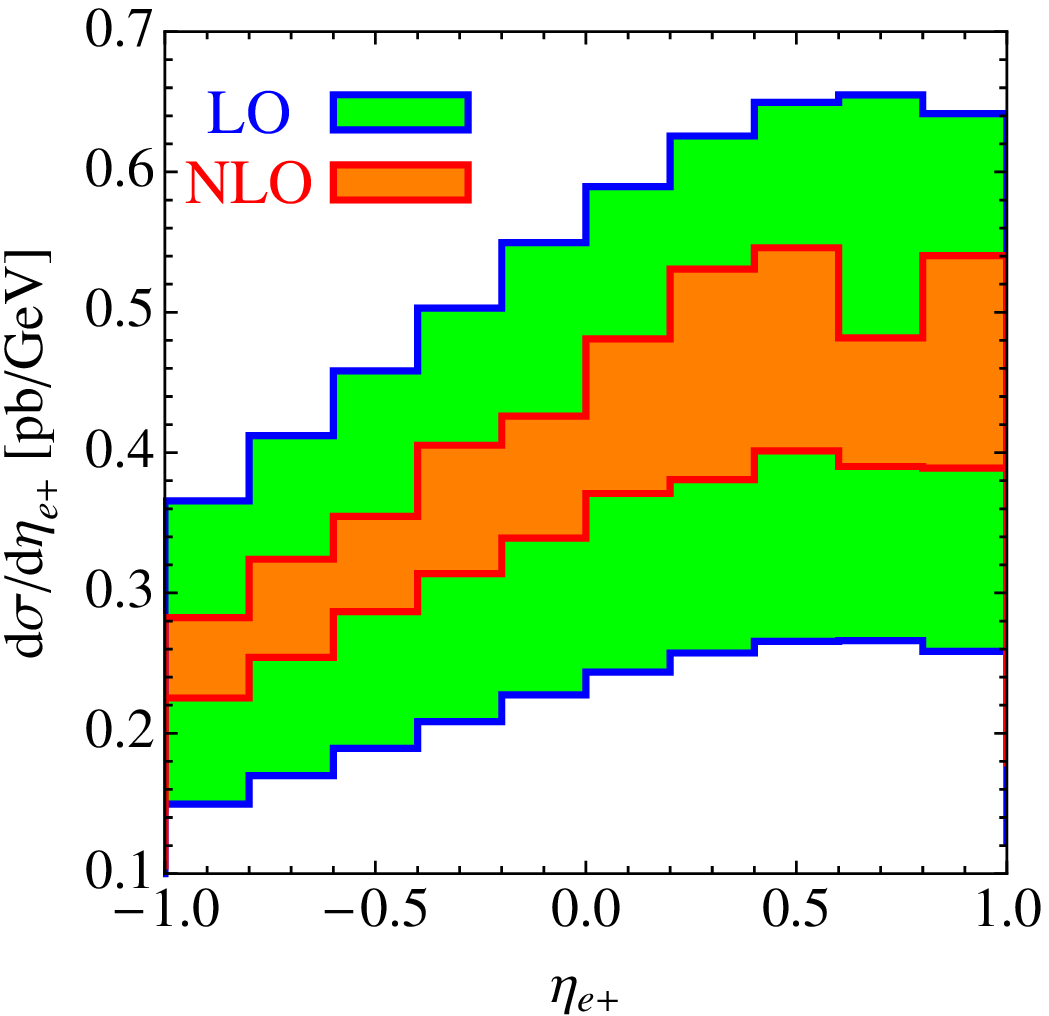}
\hspace{1cm}
\includegraphics[angle=0,scale=0.6]{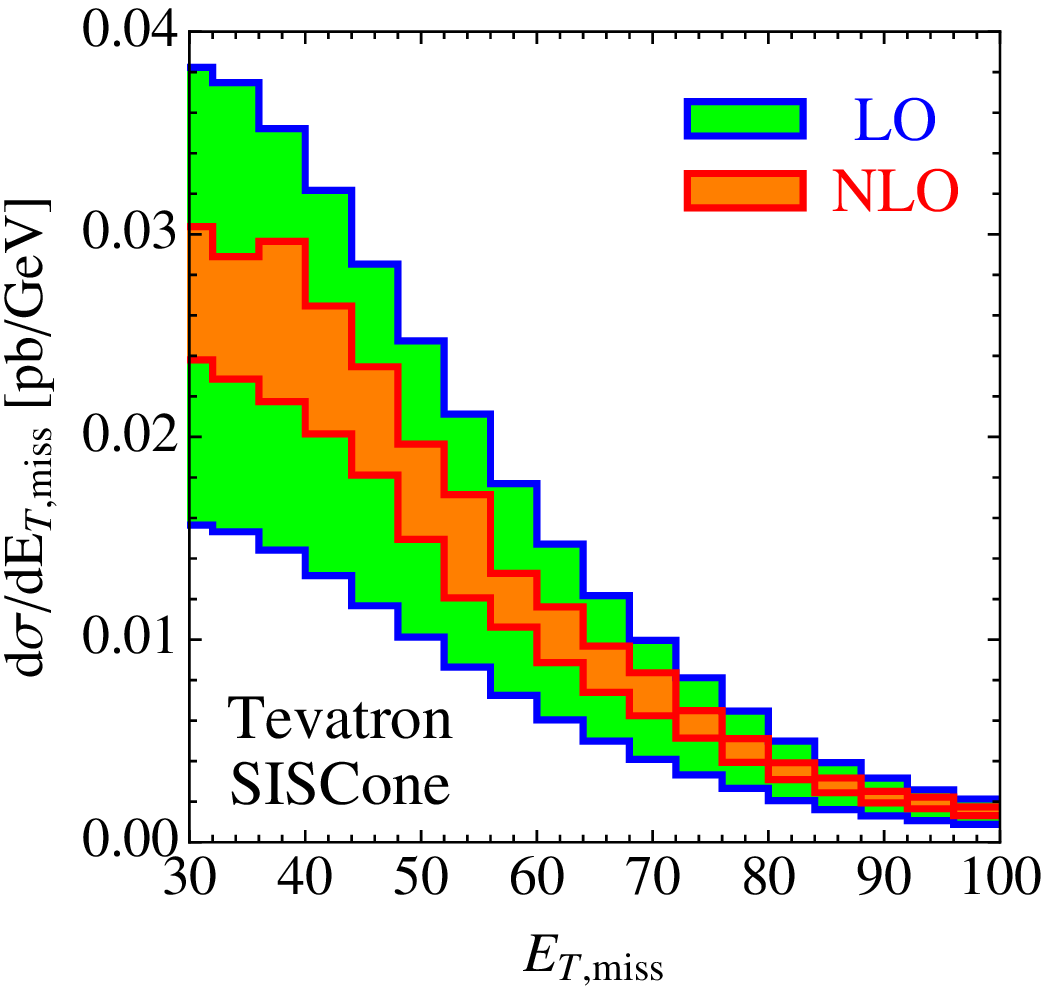}
\caption{The $e^+$ rapidity distribution and the missing energy
distribution in the $W^++ \ge 3~{\rm jet}$ sample 
using SIScone jet algorithm. All cuts and parameters are
described in the text. Bands illustrate scale
dependence at leading (green) and next-to-leading order (red).
Leading color adjustment  is applied.}
\label{fig:3}
\end{center}
\end{figure}

In Figs.~\ref{fig:2},\ref{fig:3} we present other kinematic
distributions computed through next-to-leading order. In
Fig.~\ref{fig:2}, the transverse energy distributions of hardest and
next-to-hardest jets are shown.  These distributions exhibit a shape
change similar to the shape change that we observed in the transverse
momentum distribution of the third hardest jet.

In Fig.~\ref{fig:3} the impact of NLO QCD corrections on leptonic
observables in the case of $W^+ +\ge 3$ jet production is shown. In
this case significant shape changes in both lepton rapidity
distribution and missing energy distribution do not occur, so
simulations based on leading order matrix elements should give
reliable results for the shapes.

Finally, we point out that the discussion in this Section applies to
the {\it inclusive} $W+3$ jets cross-section. In particular, the
observation that the choice of renormalization and factorization scale
$\mu = \mu_0$ leads to small corrections applies to that
observable. It is interesting to point out that the same scale choice
$\mu = \mu_0$ also works very well for {\it exclusive} $W+3$ jet
production cross-section. In that case, for $\mu = \mu_0$, the NLO QCD
corrections increase the leading order result by only about six
percent, if jets are defined with SIScone algorithm.

To conclude this Section, we compare our findings with that of Berger
et al.~\cite{BH}. As we explained in the Introduction, the computation
reported in this paper and in Ref.~\cite{BH} are not identical so full
agreement should not be expected. Nevertheless, the agreement is quite
good. For example the leading color SIScone cross-section $\sigma_{\rm
NLO}^{W+ \ge 3j}(E_\perp^{3{\rm rd~jet}}> 25~{\rm GeV}) =
0.908^{+0.044}_{-0.142}~{\rm pb}$ was reported~\cite{BH}.  Based on
the evidence from $W+1$ and $W+2$ jets, it was argued in \cite{BH}
that their leading color cross-section is within three percent of the
full color result\footnote{This claim is further supported by the
preliminary full color NLO QCD cross-section for $W+3$ jets reported
in~\cite{bhtalks}.}.  This result compares very well with our {\it
estimate} of the full color result shown in Eq.~(\ref{ourfc}).

\section{Conclusions}

We described the computation of NLO QCD radiative corrections to $W+3$
jet cross-section at the Tevatron, in the leading color
approximation. We compared our results with experimental data and
found agreement both for the total cross-section and available
differential distributions.  We point out that the scale choice $\mu =
\sqrt{p_{\perp,W}^2 + m_W^2}$ is fortunate at the Tevatron 
since for this dynamical  
scale
QCD corrections to total cross-sections are relatively small. On the
other hand, this scale choice is not a perfect solution as evident
from the fact that at NLO shapes of some distributions change.

Results presented here suggest that, after a leading color adjustment,
leading color NLO calculations are an excellent approximation to the
full color result, to within a few percent. This is more than enough
to match the experimental accuracies of Tevatron and LHC 
measurements of multi-particle final state events. From a theoretical
point of view, missing higher order (NNLO) corrections, estimated
through the residual scale dependence of the NLO result, limited
knowledge of the underlying event, and poor description of
hadronization effects give rise to much larger theoretical
uncertainties than the error due to the adjusted leading color
approximation.

Finally, we emphasize that for meaningful comparison of experimental
and theoretical results, it is important to use identical jet
algorithms.  If this is not done, systematic differences 
between different jet algorithms at the level
of ten percent or larger  can not be excluded
and indeed do occur as follows from the comparison of NLO predictions
for $W+3$ jets obtained with SIScone and anti-$k_\perp$ algorithms.
We point out that {\it any} jet algorithm can be used for a
theoretical computation as long as it is infrared safe. Unfortunately,
the JETCLU algorithm used by CDF in the published $W+3$ jet analysis
is not infrared safe and we had to switch to other jet
algorithms.  We found that NLO QCD prediction for $W+3$ jets computed
both with SIScone and anti-$k_\perp$ jet algorithms work reasonably
well in that they show agreement with data within the quoted
experimental and theoretical errors. It is important to stress,
however, that it is much better to use identical infrared safe jet
algorithms to avoid this issue in future comparisons.

\vspace*{0.2cm}
{\bf Acknowledgments}
We are grateful to Carola Berger, Fernando Febres Cordero, Lance
Dixon, and Zoltan Kunszt for useful discussions and information. We
also thank John Campbell for providing the code needed to compute
$W+2~$jets in the leading color approximation and Gavin Salam for
extensive discussions about jet algorithms.
The research of K.M. is supported by the startup package provided by
Johns Hopkins University.  G.Z. is supported by the British Science
and Technology Facilities Council.  Fermilab is operated by Fermi
Research Alliance, LLC under Contract No. DE-AC02-07CH11359 with the
United States Department of Energy.

\end{document}